\documentclass[prl,aps,graphics,floats,showpacs,preprint]{revtex4}
\usepackage{graphicx}
\usepackage{amssymb}
\usepackage{amsmath}

\begin{document}

\draft

\title{Improved time-resolved magneto-optical Kerr effect technique and dynamic magnetization reversal mechanism of
perpendicularly magnetized $L1_{\mathrm{0}}$ FePt films}

\author{X. D. Liu$^{1}$, Z. Xu$^{2}$, R. X. Gao$^{1}$, Z. F. Chen$^{1}$, T. S. Lai$^{1*}$$\footnotetext{${}^{*}${Electronic mail:
stslts@mail.sysu.edu.cn;shimingzhou@yahoo.com}}$, J. Du$^{3}$, and
S. M. Zhou$^{2}$$^{*}$}

\affiliation{$^{1}$State-Key Laboratory of Optoelectronic
Materials and Technologies, and Department of Physics, Zhongshan
(Sun Yat-Sen) University, Guangzhou 510275, China\\
\indent $^{2}$Department~of~Physics, Fudan University, Shanghai
200433,
China\\
\indent  $^{3}$National Laboratory of Solid State Microstructures,
Nanjing University, Nanjing 210093, China}

\date{\today}

\begin{abstract}

\indent The dynamic coercivity cannot be measured rigorously by
the conventional time-resolved magneto-optical Kerr effect
technique because the irreversible deviation of the transient
magnetization is accumulated. In order to remove the accumulation
effect, the alternating magnetic field is employed and
synchronized with the femtosecond laser pulse. Since the sample is
reset before each single laser pulse, the accumulation effect of
the irreversible deviation of the transient magnetization is
removed. For perpendicularly magnetized $L1_{\mathrm{0}}$ FePt
films, the dynamic magnetization reversal process is accomplished
by the nucleation of reversed domains and the pinned domain wall
motion.

\end{abstract}

\vspace{5 cm}


\maketitle


\indent  Femto-second (fs) laser induced ultrafast spin dynamics
has been studied extensively for magnetic films because of its
importance in basic research and potential applications in heat
assisted magnetic
recording~\cite{Beaurepaire1996,Hohlfeld2001,Kampen2002,Rhie2003,Vomir2005,Stanciu2007a,Stanciu2007b}.
Upon applications of fs laser, hot electrons are firstly excited.
Afterwards, spins are in non-equilibrium state because of
interactions among electrons, phonons, and spins. The transient
temperatures of electrons, spins, and phonons can be revealed by
transient reflectivity and Kerr rotation. In order to reveal the
laser-induced magnetization reversal process, dynamic hysteresis
loops were usually measured by the time-resolved magneto-optical
Kerr effect (TR-MOKE)
technique~\cite{Beaurepaire1996,Hohlfeld2001,Stanciu2007b,Wang2004,Hall2008}.
The dynamic coercivity $H_{\mathrm{C}}$ is normally much smaller
than that of the static state, that is to say, the magnetic films
become magnetically soft. In studies of CoPt$_{3}$ films, at the
delay time longer than 600 fs  $H_{\mathrm{C}}$ is equal to zero,
which was explained in terms of the paramagnetism of CoPt$_{3}$
films~\cite{Beaurepaire1998}.\\
\indent Hysteresis loops are often used to study the magnetization
reversal mechanism. The temperature should be fixed during
measurements. In the TR-MOKE measurements, however, the
temperature of the sampling location undergoes sharp rise and slow
recovery. The magnetic anisotropy constant is reduced due to the
thermal effect after the fs laser excitation. An additional
magnetic field may be induced by the linear polarized fs
laser~\cite{zhang2008}, leading to the orientation change of the
effective magnetic field. Therefore, the transient magnetization
may irreversibly deviate from the original equilibrium state when
the magnetic field is smaller than the saturation
one~\cite{Kampen2002}. As pointed out by Zhang and Roth \emph{et
al}, due to the irreversible deviation the evolution of
$H_{\mathrm{C}}$ with the delay time cannot be rigorously detected
in the conventional TR-MOKE technique~\cite{Zhang2002,Roth2008}.
To reveal the real-time evolution of the dynamic coercivity, it is
necessary to re-initialize the magnetization state after single
laser pulse shot. In this work, a photo-magnetic synchronized
TR-MOKE technique is developed based on a home-built alternating
magnetic field (AMF). As a result, the real-time
evolution of dynamic hysteresis loops is measured.\\
\indent  In our experimental setup, the AMF is generated in 2 mm
magnetic gap of an electromagnet with a nanocrystal-film-wound
core that has a response frequency of 100 kHz. The inductive
reactance of the electromagnet is cancelled by a suitable
capacitor so that the electromagnet is resonant electrically at
1.14 kHz. The electromagnet is driven by a sinusoidal current in
the amplitude of 60 A and can generate a 1.14 kHz sinusoidal AMF
with an amplitude of up to 8.0 kOe. The AMF is measured by a Hall
sensor with a response frequency of 10 kHz and sampled
synchronously. The driving power supply is externally synchronized
by fs laser pulse train from a Ti:sapphire laser amplifier with a
repetition rate tuned at 1.14 kHz, a pulse duration of 150 fs and
an energy of 0.5 mJ/pulse. A programmable electrical time delayer
with a time resolution of 25 ns is used to control the delay time
$t_{\mathrm{1}}$ between the fs pulse train and the AMF. The laser
pulses go through a standard pump-probe setup and are split into
strong pump and weak probe pulses. The pump pulses pass through an
optical delay line so that the delay time ${t_{2}}$ of the probe
pulse with respect to the pump pulse can be scanned with a fs
resolution. The pump and probe pulses are incident nearly normally
and focused to a same area on the sample. The pump spot size is at
least two times larger than the probe one, where the former one is
200 $\mu$m. The pump fluence is at least ten times more than that
of the probe. Polar Kerr rotation of the probe reflected from the
sample surface is measured by a balance optical bridge and
subsequently readout by a lock-in amplifier referenced at the
frequency of an optical
chopper which modulates the probe beam at ~340 Hz. \\
\indent It is essential to compare the difference of the
conventional and the improved TR-MOKE techniues. In the
conventional technique, the DC magnetic field is employed. On the
one hand, at every data point of the Kerr loop, the sample is
excited by multiple laser pulses. When the magnetic field is
smaller than the saturation one, the transient magnetization may
deviate from the equilibrium by each fs laser pulse, which can be
accomplished by either the precession of the transient
magnetization or the nucleation of the reversed
domains~\cite{Kampen2002,Bigot2005}. The precession may be caused
by the thermal effect on the magnetic anisotropy constant and/or
the angular momentum transfer. After the recovery procedure, the
transient magnetization finally approaches another state different
from the starting state due to the irreversibility. Between
consecutive laser pulses, irreversible deviation of the transient
magnetization is accumulated. On the other hand, the DC magnetic
field of the specific data point in the Kerr loop
changes~\emph{directly} from that of the preceding data point by a
step $\Delta H$. The starting state of a specific data point
before the laser pump is evolved from the ending state of the
preceding data point. Therefore, the deviation of the transient
magnetization is also accumulated during consecutive measurements
of subsequent data points. In a word, the so-called magnetic
softening is induced by the accumulation effect of the
irreversible deviation of the
transient magnetization.\\
\indent In order to understand the improved technique, we take the
loop at $t_{\mathrm{2}}=200$ picosecond (ps) as an example, as
shown in Fig.~\ref{Fig1}. At the data point "A" of the Kerr loop,
multiple laser pulses are applied. For each single laser pulse,
one period of the alternating magnetic field is swept. At
$t_{\mathrm{1}}=0.125$ ms, where $H_{\mathrm{A}}=2.0$ kOe, a
single pump pulse is applied. After the time delay of
$t_{\mathrm{2}}=200$ ps, the Kerr signal of the probe laser is
detected. Although the magnetic field still changes during the
pump-probe measurements, the magnetic field is almost constant
during the detection of the Kerr rotation from the probe pulse
since the range of $t_{\mathrm{2}}$ is at least six orders smaller
than the period ($T$) of $t_{\mathrm{1}}$.  It should bee pointed
out that at the beginning of each period, the sample is saturated
and that at the data point "A", \emph{ before excitation of each
laser pulse the sample is reset to the corresponding state of the
static Kerr loop without laser pump}. Therefore, the accumulation
effect is removed from either the preceding laser pulse or the
preceding data point. Above procedure is repeated to
obtain the dynamic Kerr loop by scanning $t_{\mathrm{1}}$ from 0 to $T$.\\
\indent It is significant to compare the dynamic hysteresis loops
measured by the two techniques on a 6 nm thick perpendicularly
magnetized $L1_{\mathrm{0}}$
Fe$_{\mathrm{0.5}}$Pt$_{\mathrm{0.5}}$ film, as shown in
Fig.~\ref{Fig2}. At first, the static coercivity is 2.9 kOe and
3.3 kOe in the Kerr loops measured with the dc and alternating
magnetic fields as a result of the thermal activation effect,
respectively~\cite{Suen1999,Lee1999}. Secondly, with either
approach the saturated Kerr rotation initially decreases sharply
and then recovers slowly with increasing $t_{\mathrm{2}}$.
Thirdly, as shown in Figs.~\ref{Fig2}(b)-~\ref{Fig2}(e),
$H_{\mathrm{C}}$ measured by the conventional approach is reduced
seriously by the accumulation of the irreversibly deviation of the
transient magnetization, as analyzed above. The irreversibility of
the deviation is also verified from the Kerr loops at
$t_{\mathrm{2}}=-5.0$ ps, or 877 $\mu s$ where although the
saturated transient magnetization is almost completely recovered,
$H_{\mathrm{C}}$ is still equal to zero. As shown in
Figs.~\ref{Fig2}(f)-~\ref{Fig2}(j), however,  $H_{\mathrm{C}}$
observed by the improved technique does not decrease much with
$t_{\mathrm{2}}$. At small and large $t_{\mathrm{2}}$, the dynamic
Kerr loops are almost squared and the dynamic coercivity is almost
the same as the static value. At the intermediate $t_{\mathrm{2}}$
of 600 ps, the dynamic Kerr loop is slightly slanted. In
particular, at $t_{\mathrm{2}}=200$ ps the dynamic coercivity is
smaller than the static value, as shown in Fig.~\ref{Fig1}(a). \\
\indent More importantly, with the improved method the evolution
of the dynamic Kerr loops with $t_{\mathrm{2}}$ can better reflect
the dynamic magnetization reversal process.  At small negative
magnetic fields, the magnetization reversal process is induced by
the nucleation of reversed domains instead of the
precession~\cite{Bigot2005}. This is because as shown in
Fig.~\ref{Fig3}, the transient Kerr rotation does not oscillate as
a function of $t_{\mathrm{2}}$ when the external magnetic field is
smaller than the saturation field. The nucleation process is in
turn determined by the external magnetic field, the nucleation
field, and the delay time $t_{\mathrm{2}}$. After the fs laser
excitation, the temperatures of the lattice and the spins are
raised, thereby leading a reduction of the nucleation field. At
$t_{\mathrm{2}}$ smaller than 50 ps, however, the nucleation of
the reversed domains cannot happen due to its large timescale. At
$t_{\mathrm{2}}$ of hundreds of picoseconds, the nucleation effect
becomes more prominent. Since the large negative magnetic field
favors to induce the nucleation, the reduction of the Kerr
rotation is increased at more negative magnetic field. At
$t_{\mathrm{2}}$ of nanoseconds, the temperatures of the lattice
and the spins recover and the nucleation field is increased,
thereby weakening the nucleation effect. Accordingly, the dynamic
Kerr loops are squared at small and large $t_{\mathrm{2}}$ but
slanted at intermediate $t_{\mathrm{2}}$. Moreover, at all
$t_{\mathrm{2}}$ the sharp reduction of the transient Kerr
rotation occurs at the switching field of the static Kerr loop as
marked by the dot line in Fig.~\ref{Fig2}. For the FePt film, the
static magnetization reversal process is accompanied by the pinned
domain wall motion (not shown). Although the pinning field may be
reduced, the variation of the switching field cannot be observed
in the $t_{\mathrm{2}}$ region of several hundreds of picoseconds
because the pinned domain wall motion happens in the timescale of
nanoseconds. More remarkably, although the transient magnetization
is mostly reversed as demonstrated by the negative transient Kerr
rotations near $t_{\mathrm{2}}=400$ ps when the magnetic field,
such as -3.0 kOe and -3.4 kOe, is smaller than the switching
field, it is still recovered, as shown in Fig.~\ref{Fig3}.
Finally, it should be pointed out that $H_{\mathrm{C}}$ observed
by the conventional TR-MOKE technique also strongly depends on the
pump fluence. At small fluences, it may be equal to the static
value. This is because the additional magnetic field induced by
the linearly polarized laser and the thermal effect become small,
leading to a negligible deviation of the transient magnetization
from the equilibrium position and the
reduction of  $H_{\mathrm{C}}$ is suppressed.\\
\indent In conclusion, in order to measure the dynamic coercivity
rigorously, the conventional TR-MOKE technique is improved. The fs
laser pump and the alternating magnetic field are synchronized, in
which the magnetic field is controlled by modifying
$t_{\mathrm{1}}$. Before each fs laser pulse, the sample is reset
magnetically, thereby leading to the elimination of the
accumulation effect of the irreversible deviation of the transient
magnetization. With the improved TR-MOKE technique, the dynamic
magnetization reversal process of the perpendicularly magnetized
$L1_{\mathrm{0}}$ FePt film is better revealed. It consists of the
nucleation of reversed domains at small negative magnetic fields
and then by the pinned domain wall motion at the switching field.
The dynamic coercivity changes non-monotonically as a function of
$t_{\mathrm{2}}$ with a minimum at intermediate $t_{\mathrm{2}}$
whereas the switching field almost does not change. \\
\indent This work is supported partially by Natural Science
Foundation of China under grant Nos.50625102, 50871030, 10874076,
60678009, 60490290, and 10874247, the National Basic Research
Program of China under grant Nos. 2007CB925104 and 2009CB929201,
973-project under grant No. 2006CB921300.

\begin{figure}[p]
\begin{center}
FIGURE CAPTIONS
\end{center}

\flushleft \indent Figure 1  Typical dynamic polar Kerr loop
measured by the improved TR-MOKE (a), and schematic pictures of AC
magnetic field within one period (b) and the delay time between
the pump and probe pulses (c). The data point "A" in (a)
corresponds to that of ( $t_{\mathrm{A}}$, $H_{\mathrm{A}}$) in
(b). In (b) and (c), the time relationship between the alternating
magnetic field and the pump pulse, and that between the pump and
probe
pulses are demonstrated. The scale of $t_{\mathrm{2}}$ is exaggerated greatly for clarity\\

\indent Figure 2 Polar dynamic Kerr loops measured by the
conventional (left column) and
 the improved (right column) TR-MOKE techniques with 0.5 ps (b, g) and 50 ps (c, h) and 600 ps (d, i) and
-5 ps, i.e., 877 $\mu$s (e, j). In comparison, the Kerr loops
measured by DC (a) and alternating (f) magnetic fields without
pump fluence are given. The inset numbers refers to the delay time
$t_{\mathrm{2}}$. $\theta_{\mathrm{KS0}}$ refers to the corresponding saturation Kerr rotation at the static state.\\

\indent Figure 3 Transient polar Kerr rotation versus the delay
time $t_{\mathrm{2}}$ at negative magnetic fields. \\

\end{figure}

\begin{figure}[p]
\begin{center}
\resizebox*{6 in}{!}{\includegraphics*{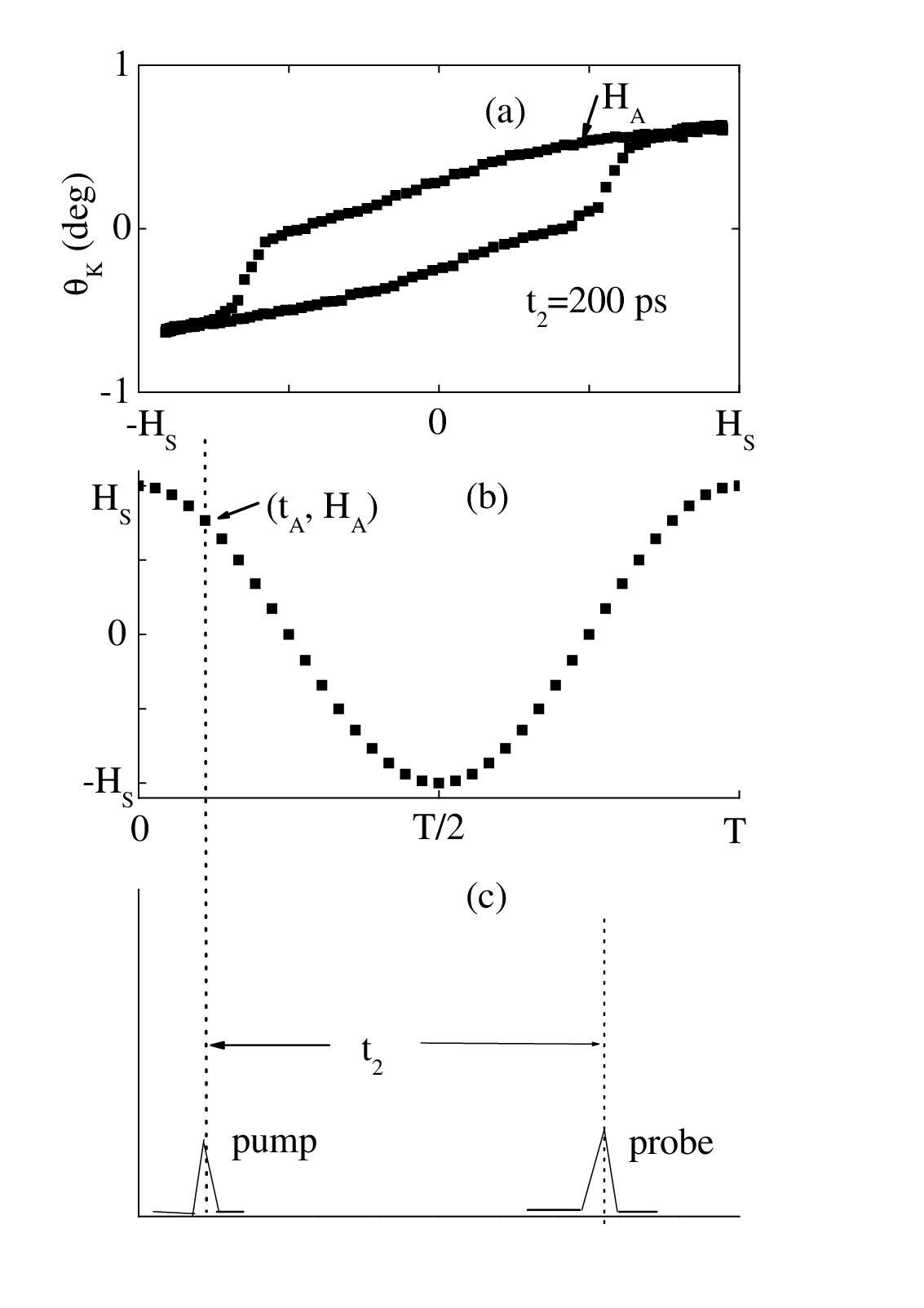}} \caption{}
\label{Fig1}
\end{center}
\end{figure}

\begin{figure}[p]
\begin{center}
\resizebox*{6  in}{!}{\includegraphics*{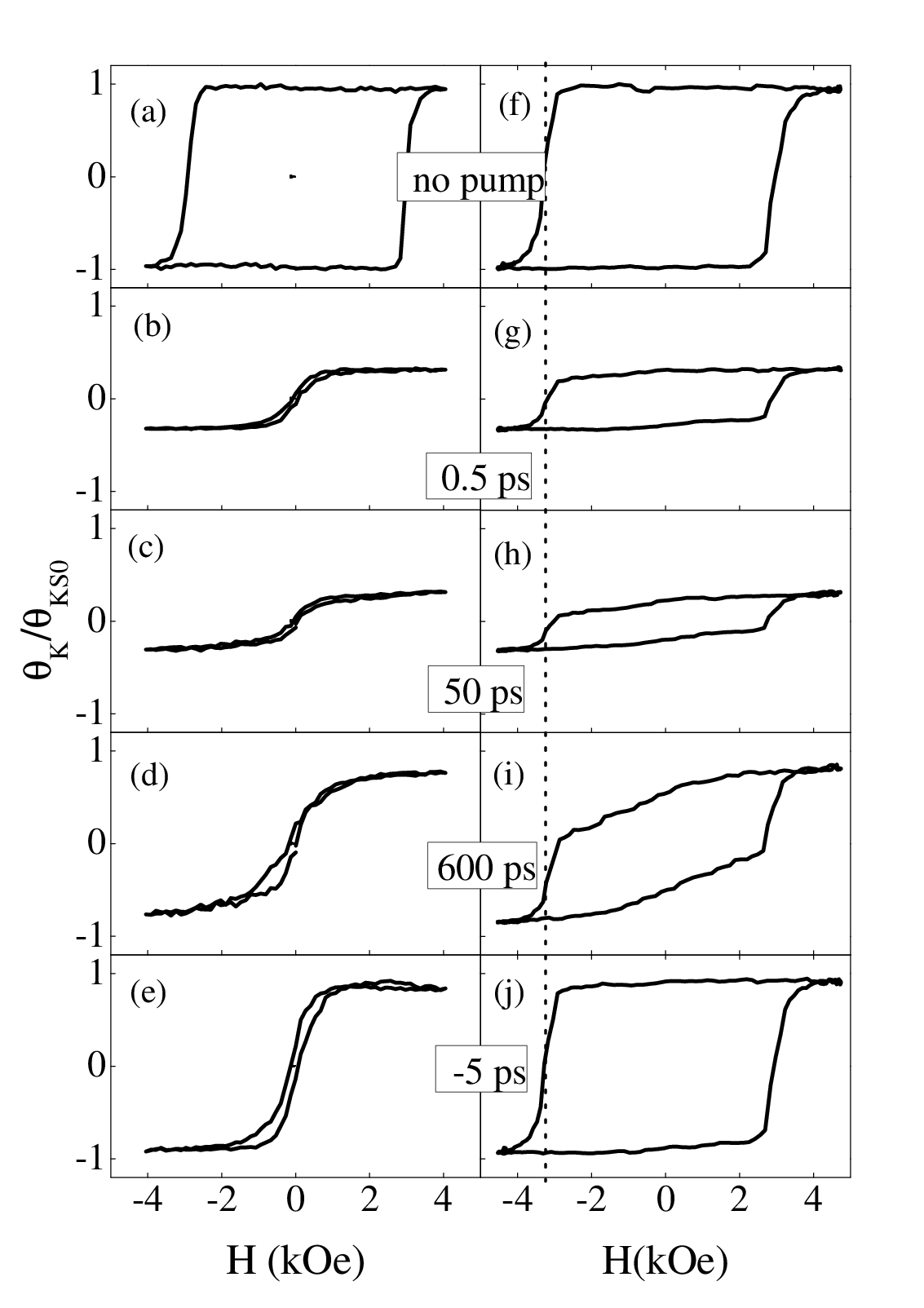}}
\end{center}
\caption{} \label{Fig2}
\end{figure}

\begin{figure}[p]
\begin{center}
\resizebox*{6 in}{!}{\includegraphics*{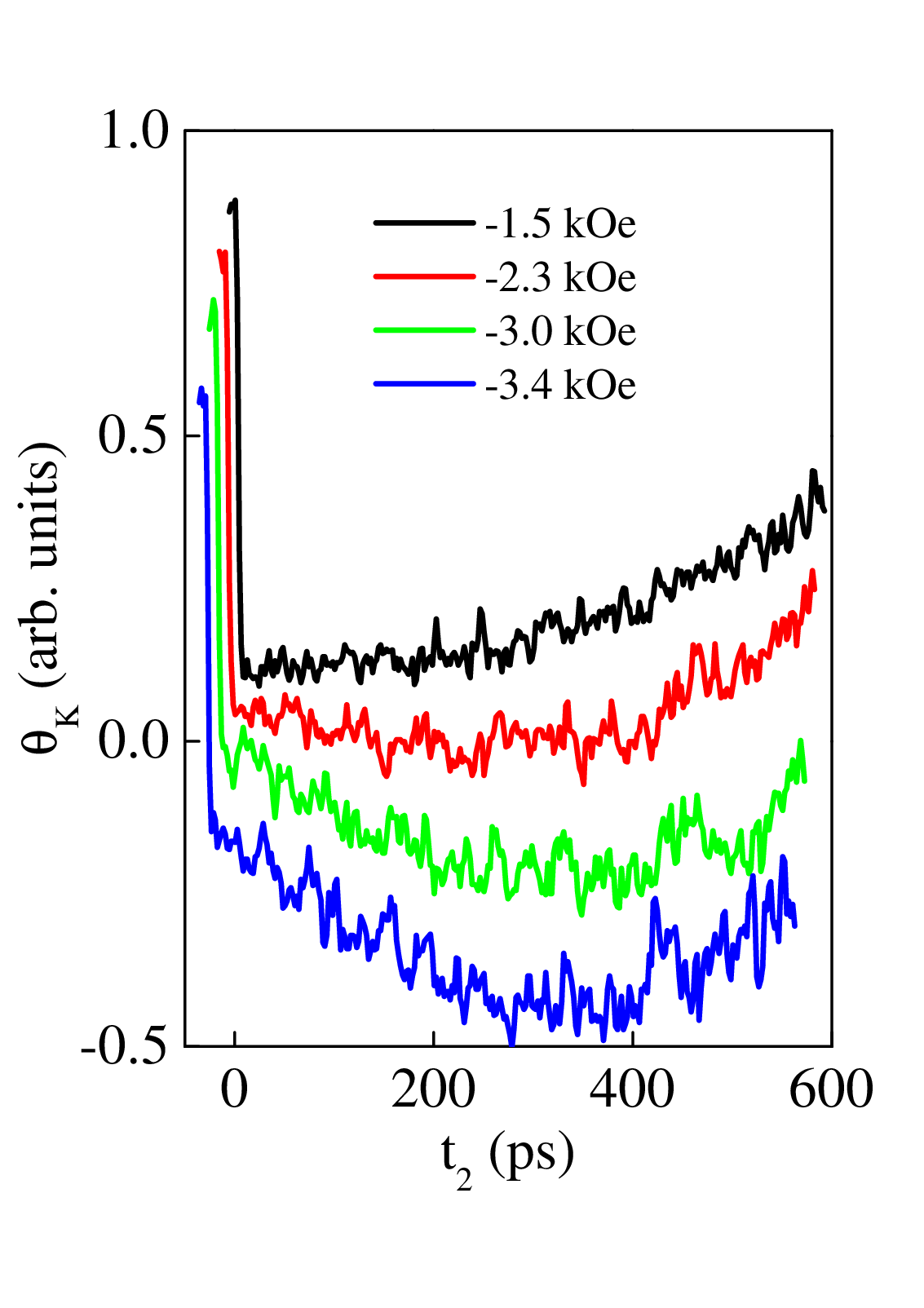}}
\end{center}
\caption{} \label{Fig3}
\end{figure}

\end{document}